\pdfoutput=1
\documentclass[showpacs,superscriptaddress,aps,pra, twocolumn]{revtex4-1}
\usepackage[utf8]{inputenc}

\usepackage[a4paper,top=3cm,bottom=2cm,left=3cm,right=3cm,marginparwidth=2cm]{geometry}

\usepackage{physics}
\usepackage{amsmath}
\usepackage{amsfonts}
\usepackage{graphicx}
\usepackage{url}
\usepackage[colorinlistoftodos]{todonotes}
\usepackage[colorlinks=true, allcolors=blue]{hyperref}
\usepackage{subcaption}
\def\p{\mathbf{p}}

\def\q{\mathbf{q}}
\usetikzlibrary{arrows}
\usepackage{bm}
\usepackage{bbm}
\usepackage{amsthm}
\def\RR{\mathbbm{R}}
\def\st{\textrm{subject to }}
\newtheorem{thm}{Theorem}

\usepackage[T1]{fontenc}

\numberwithin{equation}{section}
\begin{document}
\title{From  nonlocality quantifiers for behaviors to nonlocality quantifiers for states}
\author{Ari Patrick}
\affiliation{Department of Mathematical Physics, Institute of Physics, University of S\~ao Paulo, R. do Mat\~ao 1371, S\~ao Paulo 05508-090, SP, Brazil}

\author{Giulio Camillo}
\affiliation{Department of Mathematical Physics, Institute of Physics, University of S\~ao Paulo, R. do Mat\~ao 1371, S\~ao Paulo 05508-090, SP, Brazil}

\author{Fernando Parisio}
\affiliation{Departamento de Física, Universidade Federal de Pernambuco, Recife, Pernambuco 50670-901 Brazil}

\author{Barbara Amaral}
\affiliation{Department of Mathematical Physics, Institute of Physics, University of S\~ao Paulo, R. do Mat\~ao 1371, S\~ao Paulo 05508-090, SP, Brazil}

\date{June 2022}

\begin{abstract}
  We define an alternative way of quantifying nonlocality of states based on Bell nonlocality of behaviors, called the \emph{trace-weighted nonlocal volume}. The construction is based on the nonlocal volume,  a quantifier of nonlocality for states that counts the volume of the set of measurements that give rise to nonlocal behaviors when applied to this state, plus the   trace distance, a quantifier of nonlocality for behaviors based on the distance between the behavior and  the local set. The key difference from preceding candidates was the introduction of a quantifier of nonlocality to weight each contribution from behaviors in the nonlocal volume. We list some interesting properties of this quantifier and investigate the  $(2,2,2)$ and $(2,3,2)$ scenarios. We show that the weak anomaly of nonlocality for the $(2,2,3)$ scenario persists, but the local minimum for nonlocality with the trace-weighted nonlocal volume occurs in a different state as compared to the minimum for the non-weighted version, showing that the weak anomaly is not an intrinsic characteristic of the scenario, but is  dependent of the choice of quantifier. 
\end{abstract}

\maketitle

\section{Introduction}

Quantum nonlocality is an eccentric manifestation of quantum predictions when viewed through the eyes of a classical observer. Bell nonlocality is a consequence of entanglement, and implies that the statistics of certain measurements performed on spatially separated entangled quantum systems can not be explained by local hidden variable models \cite{bell1964einstein}. The debate around quantum nonlocality started during the development of the mathematical foundations of quantum theory  \cite{einstein1935can} and gained the status of a theorem mainly with the works of Bell  \cite{bell1964einstein}. Besides its importance in foundations of quantum theory, quantum nonlocality has been raised to the status of a \emph{physical resource} due to its intimate relationship with quantum information science, where it is necessary to enhance our computational and information processing capabilities in a variety of applications \cite{RevModPhys.82.665, RevModPhys.86.419}.

Since nonlocality has been identified as a resource, it is essential to formulate  \emph{resource theories of nonlocality}, not only allowing for operational interpretations but as well for the precise quantification of nonlocality. A resource theory provides a robust structure for the  treatment of a property as a physical resource, allowing for its characterization, quantification and manipulation \cite{amaral2019resource}. A resource theory  consists of three essential ingredients: a set of \emph{objects}, which represent the physical entities which can contain the resource, and a subset of objects called \emph{free objects}, which are the objects that do not contain the resource; a special class of transformations, called \emph{free operations}, which fulfill the essential requirement of mapping every free object of the theory into a free object; and finally, a \emph{quantifier}, which maps each object to a real number that represents quantitatively how much resource this object contains, and which is monotonic under the action of free operations.

Given its pervasive importance, the resource theory of entanglement \cite{Horodecki2009} is arguably the most well understood and vastly explored and thus has become the archetypal example for the development of resource theories of other quantum phenomena \cite{Gour2008,Brandao2013,Vicente2014,Winter2016,Coecke2016,Abramsky2017,Amaral2017}. Entanglement is a necessary ingredient for Bell nonlocality, but despite  their close connection,  entanglement and nonlocality refer to truly different resources \cite{Brunner2014}, as  there are entangled states that can only give rise to local correlations \cite{Werner1989}. Therefore it is important to  develop resource theories of nonlocality that are independent of entanglement.

We will consider a bipartite Bell scenario  where two  distant parts, Alice and Bob, share a possibly correlated pair of physical systems, in which they perform measurements, obtaining measurement outcomes. Alice and Bob are not allowed to communicate and the choices of inputs  are assumed to be independent.  In the discussion of nonlocality, it is generally   assumed that these physical systems are described by closed boxes and that Alice and Bob do not have access to the details of the state of the system and the measurements being performed. Hence, the description of Alice and Bob's outcomes is given by a set of probability distributions, called \emph{behavior}, that depends on the physical assumptions we make about the systems they share. In this situation, we are interested in a resource theory where the set of objects is the set of behaviors, and the set of free objects is the set of local behaviors, those consistent with a description through a local hidden variable model.

It is also interesting to focus on the state shared by Alice and Bob and to study how nonlocal behaviors can be generated using this state. This is intimately related to the idea of how much nonlocality can be ``extracted'' from the state. 
 In this situation, we are interested in a resource theory where the set of objects is the set of quantum states, perhaps with a fixed dimension, and the set of free objects is the set of states for which every possible set of measurements performed by Alice and Bob generates local behaviors. In this contribution we investigate how nonlocality quantifiers for behaviors can be used to create nonlocality quantifiers for states. Moreover, we investigate the phenomenon known as \emph{anomaly of nonlocality} \cite{PhysRevA.65.052325}, with special attention to the \emph{weak anomaly} observed in \cite{PhysRevA.96.012101}.

 Recently,  several  nonlocality quantifiers for behaviors  have been proposed \cite{Brunner2014,Popescu1994,Eberhard1993,Toner03,Pironio2003,van2005,Acin2005,Junge2010,Hall2011,Chaves2012,Vicente2014,Fonseca2015,Rosier2017,Chaves2015b,Ringbauer2016,Montina2016,Brask2017,GA17}, and  a proper resource theory of nonlocality for behaviors has been developed \cite{Vicente2014,GA17}, thus allowing for a formal proof that previously introduced quantities indeed provide good measures of nonlocality. Of special importance to this work is the nonlocality quantifier based on the trace  distance  presented in reference \cite{PhysRevA.97.022111}, which measures the $l_1$-distance between a given behavior and the local polytope. It is shown in this reference that, for the $ (2,2,d) $ scenario, the maximum numerical violation of the CGLMP inequality grows with $d$, and likewise, the maximum value of the trace distance follows the same trend.

We also have several examples of nonlocality quantifiers for a resource theory where the objects are  quantum states  \cite{PhysRevA.65.052325, Eberhard1993,  Acin2005, Brunner_2005, Fonseca2015, PhysRevA.98.022132}. Different measures will have different operational meanings and do not necessarily have to agree on the ordering for the amount of nonlocality. For instance, a natural way of quantifying nonlocality is the maximum violation of a specific Bell inequality allowed by a given quantum state. However, we might also be interested in quantifying the nonlocality of a state by the amount of noise (e.g., detection inefficiencies) it can stand before becoming local  \cite{PhysRevLett.85.4418, Laskowski_2014, PhysRevA.65.052325}. Interestingly, these two measures can be inversely related as demonstrated by the fact that in the CHSH scenario \cite{Clauser1969} the resistance against detection inefficiency increases as the  entanglement of the state decreases \cite{Eberhard1993}, also reducing the violation the of CHSH inequality.

 A curious feature of quantifiers based on Bell inequalities is that the highest violation value is not always  achieved by the maximum entangled state, a property known  as \emph{anomaly of nonlocality} \cite{PhysRevA.65.052325}. This is a counter-intuitive property since it was initially expected that ``more'' entanglement would lead to ``more'' nonlocality.

This observation led to the investigation of whether the anomaly was a property of this specific quantifier only, or if it was  a proof that nonlocality and entanglement are indeed distinct  resources. With this purpose in mind, the authors of \cite{Fonseca2015}  proposed a nonlocality quantifier called the \emph{volume of violation}, that counts the volume of the set of measurements that give rise to violation of the chosen inequality when applied to this state.
This quantifier solves the anomaly for qutrits, since in this case maximally entangled qutrits are maximally nonlocal (with respect to the volume of violation). In reference  \cite{PhysRevA.96.012101} the authors analyse several scenarios, including up to 5 qubits and different entangled states. They show that the probability of violation increases as we increase the number of measurements and the maximum violation for these dimensions is obtained by the maximally entangled state. Several properties of these quantifiers were proven in reference \cite{lipinska2018towards}.

The disadvantage  of the volume of violation is that it focuses in only one inequality, and for scenarios with more than two parties, measurements or outcomes it is known that there are many nonequivalent families of Bell inequalities. Hence, a single inequality can not capture the entire structure of the set of local behaviors. This implies that the volume of violation does not take into account a large portion of nonlocal behaviors that can be generated with the state, and this can lead to many undesired properties. For example, considering only a single CGLMP inequality, the anomaly of nonlocality reappears for $d>7$ \cite{PhysRevA.98.022132}.

Another nonlocality quantifier was proposed to solve this problem. The definition is similar to the definition of the volume of violation, but it takes the entire space of  behaviors into consideration: it counts the volume of the set of measurements that give rise to nonlocal behaviors when applied to this state. The maximum nonlocal volume is obtained with the maximum entangled state for dimensions up to $d=10$ \cite{PhysRevA.98.042105}, which  indicates that the quantifiers based on the volume has a greater relevance when calculated over the polytopes than considering a specific inequality. Recently, a quantifier based on both the volume of violation and noise resistance was studied, where these two quantities are combined \cite{PhysRevA.101.012116}, and an experiment based on this combination has also been performed \cite{Barasinski2021experimentally}.

However, the nonlocal volume does not solve the anomaly of nonlocality completely. It was shown in reference \cite{PhysRevA.96.012101} that this quantifier exhibits a \emph{weak anomaly} for qutrits: the maximum volume is achieved for the maximum entangled state, but the nonlocal volume is not a monotonic function of the state's entanglement. We suggest that the appearance of this non-monotonicity, despite the coincidence of the maxima, is an unavoidable manifestation of the intrinsic inequivalence between entanglement and nonlocality.

In this work, we use the nonlocal volume plus the   trace distance \cite{PhysRevA.97.022111} to define a new nonlocality quantifier for states, called the \emph{trace-weighted nonlocal volume}. The proposal is to use the numerical value of the trace distance to the set of local behaviors as a weight in the definition of the nonlocal volume. Initially, we reproduce some results in previous works for the scenarios $(2,2,2)$ and $(2,3,2)$. We also  verify if this new measure of nonlocality satisfies the properties listed in \cite{lipinska2018towards}. Finally, we use our quantifier to study the weak anomaly of nonlocality reported in \cite{PhysRevA.96.012101}.

This work is organized as follows: In sec. \ref{section2}, we describe the scenario of interest. In sec. \ref{section3} we
define a quantifier for nonlocality of behaviors based on the trace distance. In sequence,  sec. \ref{section4}  discusses  the volume of violation, the results obtained for this quantifier, in addition to its generalization for the space of behaviors, called the nonlocal volume. Finally, we propose a new quantifier similar to the nonlocal volume, where we use the trace distance as weight for the calculation of the integral. In the sec.\ref{section5} we show that our quantifier satisfies some properties that are expected by any nonlocality quantifier for states. We then investigate  some  Bell
scenarios in sec. \ref{section6}. In sec. \ref{section7} we deal with the weak anomaly from the perspective of our quantifier. Finally, in sec. \ref{section8}, we give some of the most relevant conclusions of the work.

\section{The geometry of the set of Bell correlations}\label{section2}

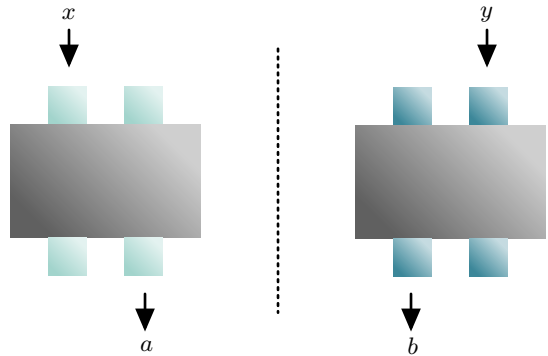
\begin{figure}
    \centering
   \definecolor{ududff}{rgb}{0.6372549019607844,0.7313725490196079,0.8019607843137255}
\definecolor{ccqqqq}{rgb}{0.6372549019607844,0.8313725490196079,0.8019607843137255}
\definecolor{wqwqwq}{rgb}{0.3764705882352941,0.3764705882352941,0.3764705882352941}
\definecolor{qqwuqq}{rgb}{0.2372549019607844,0.5313725490196079,0.6019607843137255}
\begin{tikzpicture}[scale=0.5,line cap=round,line join=round,>=triangle 45,x=1cm,y=1cm]
\fill[line width=2pt,color=qqwuqq,fill=qqwuqq,fill opacity=1,shading = rectangle, left color=qqwuqq, right color=qqwuqq!30!white, shading angle=135] (2.04,4.96) -- (2.04,3.96) -- (3.04,3.96) -- (3.04,4.96) -- cycle;
\fill[line width=2pt,color=qqwuqq,fill=qqwuqq,fill opacity=1,shading = rectangle, left color=qqwuqq, right color=qqwuqq!30!white, shading angle=135] (4.04,4.96) -- (4.04,3.96) -- (5.04,3.96) -- (5.04,4.96) -- cycle;
\fill[line width=2pt,color=wqwqwq,fill=wqwqwq,fill opacity=1,shading = rectangle, left color=wqwqwq, right color=wqwqwq!30!white, shading angle=135] (1.04,3.96) -- (1.04,0.96) -- (6.04,0.96) -- (6.04,3.96) -- cycle;
\fill[line width=2pt,color=qqwuqq,fill=qqwuqq,fill opacity=1,shading = rectangle, left color=qqwuqq, right color=qqwuqq!30!white, shading angle=135] (2.04,0.96) -- (2.04,-0.04) -- (3.04,-0.04) -- (3.04,0.96) -- cycle;
\fill[line width=2pt,color=qqwuqq,fill=qqwuqq,fill opacity=1,shading = rectangle, left color=qqwuqq, right color=qqwuqq!30!white, shading angle=135] (4.04,0.96) -- (4.04,-0.04) -- (5.04,-0.04) -- (5.04,0.96) -- cycle;
\fill[line width=2pt,color=ccqqqq,fill=ccqqqq,fill opacity=1,shading = rectangle, left color=ccqqqq, right color=ccqqqq!30!white, shading angle=135] (-7.04,5) -- (-7.04,4) -- (-6.04,4) -- (-6.04,5) -- cycle;
\fill[line width=2pt,color=ccqqqq,fill=ccqqqq,fill opacity=1,shading = rectangle, left color=ccqqqq, right color=ccqqqq!30!white, shading angle=135] (-5.04,5) -- (-5.04,4) -- (-4.04,4) -- (-4.04,5) -- cycle;
\fill[line width=2pt,color=wqwqwq,fill=wqwqwq,fill opacity=1,shading = rectangle, left color=wqwqwq, right color=wqwqwq!30!white, shading angle=135] (-8.04,4) -- (-8.04,1) -- (-3.04,1) -- (-3.04,4) -- cycle;
\fill[line width=2pt,color=ccqqqq,fill=ccqqqq,fill opacity=1,shading = rectangle, left color=ccqqqq, right color=ccqqqq!30!white, shading angle=135] (-7.04,1) -- (-7.04,0) -- (-6.04,0) -- (-6.04,1) -- cycle;
\fill[line width=2pt,color=ccqqqq,fill=ccqqqq,fill opacity=1,shading = rectangle, left color=ccqqqq, right color=ccqqqq!30!white, shading angle=135] (-5.04,1) -- (-5.04,0) -- (-4.04,0) -- (-4.04,1) -- cycle;
\draw [line width=1pt,dotted] (-1,6)-- (-1,-1);
\draw [->,line width=1pt] (-6.5,6.5) -- (-6.5,5.5);
\draw [->,line width=1pt] (4.5,6.5) -- (4.5,5.5);
\draw [->,line width=1pt] (2.5,-0.5) -- (2.5,-1.5);
\draw [->,line width=1pt] (-4.5,-0.5) -- (-4.5,-1.5);
\draw[color=black] (-6.505838962730431,6.924319717669096) node {$x$};
\draw[color=black] (4.487920702785661,6.924561526992208) node {$y$};
\draw[color=black] (2.5425229018918674,-1.8867046568297686) node {$b$};
\draw[color=black] (-4.45487694008271,-1.8867046568297686) node {$a$};
\end{tikzpicture}
    \caption{Alice and Bob have each a measurement device with inputs $x$ and $y$ and outputs $a$ and $b$. The device is described by the set of probability distributions $p\left(ab\vert xy\right)$, that gives the probability of outcomes $a$ and $b$ given inputs $x$ and $y$.}
    \label{fig:bipartite}
\end{figure}

We consider a bipartite Bell scenario  in which Alice and Bob perform measurements labeled by variables $A_x$ and $B_y$, obtaining measurement outcomes described by  variables $a$ and $b$, respectively, as shown in fig. \ref{fig:bipartite}. The description of Alice and Bob's outcomes is given by a set of probability distributions $p(a,b \vert x,y)$, called a \emph{behavior}, that gives the probability of outcomes $a$ and $b$ given inputs $x$ and $y$. The set of behaviors under consideration depends on the physical assumptions we make about the systems Alice and Bob share.

The fact that Alice and Bob are spatially separated and cannot communicate with each other implies that the statistics of a measurement on one part is independent of the measurement choice of the other. These assumptions imply a set of linear constraints  known as \emph{nonsignalling conditions:}
\begin{equation}\label{NS}
\begin{aligned}
    & & p(a \vert x) = \sum_{b} p(a,b \vert x,y)= \sum_{b} p(a,b \vert x,y^{\prime}) \\
& & p(b \vert y) = \sum_{a} p(a,b \vert x,y)= \sum_{a} p(a,b \vert x^{\prime},y).
\end{aligned}
\end{equation}

 A stronger constraint  on the description of the experiment is that the statistics of Alice and Bob be consistent with the assumption of local causality, which implies  that the probability distributions can be decomposed as
\begin{equation}
\label{LHV}
p(a,b \vert x,y)= \sum_{\lambda} p(\lambda) p(a \vert x,\lambda) p(b \vert y,\lambda).
\end{equation}
For this type of behavior, correlations between Alice and Bob are  mediated by the variable $\lambda$ that thus suffices to compute the probabilities of each of the outcomes, that is, $p(a\vert x,y,b,\lambda)=p(a\vert x,\lambda)$, and similarly for $b$. The behaviors that can be decomposed in this way are called \emph{local  behaviors}.

The result known as Bell's theorem \cite{bell1964einstein} states that Alice and Bob can perform measurements in a entangled quantum state to generate behaviors that cannot be decomposed in the form \eqref{LHV}. 
These can be obtained by local measurements $M^x_a$ and $M^y_b$ on distant parts of a bipartite state $\rho$ that, according to quantum theory, can be  described by
\begin{equation}
\label{quantum}
p(a,b \vert x,y)= \mathrm{Tr} \left[ \left(M^x_a \otimes M^y_b \right) \rho \right].
\end{equation}

\begin{figure}[ht]
    \centering
    \definecolor{xdxdff}{rgb}{0.49019607843137253,0.49019607843137253,1}
\definecolor{ududff}{rgb}{0.30196078431372547,0.30196078431372547,1}
\definecolor{bcduew}{rgb}{0.7372549019607844,0.8313725490196079,0.9019607843137255}
\definecolor{wqwqwq}{rgb}{0.2764705882352941,0.2764705882352941,0.2764705882352941}
\definecolor{aqaqaq}{rgb}{0.6274509803921569,0.6274509803921569,0.6274509803921569}
\definecolor{ffcydc}{rgb}{0.6372549019607844,0.7313725490196079,0.8019607843137255}
\begin{tikzpicture}[scale=0.6,line cap=round,line join=round,>=triangle 45,x=1cm,y=1cm]
\fill[line width=1.2pt,color=ffcydc,fill=ffcydc,fill opacity=1] (6,-6) -- (6,-2) -- (12,-2) -- (12,-6) -- cycle;
\fill[line width=1.2pt,color=aqaqaq,fill=aqaqaq,fill opacity=1] (6,-2) -- (9,2) -- (12,-2) -- cycle;
\draw[color=wqwqwq] (6,-2) -- (9,2);
\fill[line width=1.2pt,color=ffcydc,fill=ffcydc,fill opacity=1] (15,3) -- (15,2.5) -- (15.5,2.5) -- (15.5,3) -- cycle;
\draw[color=black] (16.5,2.75) node {$\mathcal{L}$};
\fill[line width=1.2pt,color=aqaqaq,fill=aqaqaq,fill opacity=1] (14,2) -- (14,1.5) -- (14.5,1.5) -- (14.5,2) -- cycle;
\fill[line width=1.2pt,color=ffcydc,fill=ffcydc,fill opacity=1] (15,1.5) -- (15.5,1.5) -- (15.5,2) -- (15,2) -- cycle;
\fill[line width=1.2pt,color=wqwqwq,fill=wqwqwq,fill opacity=1] (13,2) -- (13,1.5) -- (13.5,1.5) -- (13.5,2) -- cycle;
 \draw[color=black] (16.5,1.75) node {$\mathcal{NS}$};
\fill[line width=1.2pt,color=wqwqwq,fill=wqwqwq,fill opacity=1] (14,1) -- (14,0.5) -- (14.5,0.5) -- (14.5,1) -- cycle;
\fill[line width=1.2pt,color=ffcydc,fill=ffcydc,fill opacity=1] (15,1) -- (15,0.5) -- (15.5,0.5) -- (15.5,1) -- cycle;
\draw[color=black] (16.5,0.75) node {$\mathcal{Q}$};
\draw [line width=1.2pt,color=ffcydc] (6,-6)-- (6,-2);
\draw [line width=1.2pt,color=ffcydc] (6,-2)-- (12,-2);
\draw [line width=1.2pt,color=ffcydc] (12,-2)-- (12,-6);
\draw [line width=1.2pt,color=ffcydc] (12,-6)-- (6,-6);
\draw [line width=1.2pt,color=aqaqaq] (6,-2)-- (9,2);
\draw [line width=1.2pt,color=aqaqaq] (9,2)-- (12,-2);
\draw [line width=1.2pt,color=aqaqaq] (12,-2)-- (6,-2);
\draw [shift={(9,-5)},line width=1.2pt,color=wqwqwq,fill=wqwqwq,fill opacity=1]  plot[domain=0.7853981633974483:2.356194490192345,variable=\t]({1*4.242640687119286*cos(\t r)+0*4.242640687119286*sin(\t r)},{0*4.242640687119286*cos(\t r)+1*4.242640687119286*sin(\t r)});
\draw [line width=1.2pt,color=ffcydc] (15,3)-- (15,2.5);
\draw [line width=1.2pt,color=ffcydc] (15,2.5)-- (15.5,2.5);
\draw [line width=1.2pt,color=ffcydc] (15.5,2.5)-- (15.5,3);
\draw [line width=1.2pt,color=ffcydc] (15.5,3)-- (15,3);
\draw [line width=1.2pt,color=aqaqaq] (14,2)-- (14,1.5);
\draw [line width=1.2pt,color=aqaqaq] (14,1.5)-- (14.5,1.5);
\draw [line width=1.2pt,color=aqaqaq] (14.5,1.5)-- (14.5,2);
\draw [line width=1.2pt,color=aqaqaq] (14.5,2)-- (14,2);
\draw [line width=1.2pt,color=ffcydc] (15,1.5)-- (15.5,1.5);
\draw [line width=1.2pt,color=ffcydc] (15.5,1.5)-- (15.5,2);
\draw [line width=1.2pt,color=ffcydc] (15.5,2)-- (15,2);
\draw [line width=1.2pt,color=ffcydc] (15,2)-- (15,1.5);
\draw [line width=1.2pt,color=wqwqwq] (14,1)-- (14,0.5);
\draw [line width=1.2pt,color=wqwqwq] (14,0.5)-- (14.5,0.5);
\draw [line width=1.2pt,color=wqwqwq] (14.5,0.5)-- (14.5,1);
\draw [line width=1.2pt,color=wqwqwq] (14.5,1)-- (14,1);
\draw [line width=1.2pt,color=ffcydc] (15,1)-- (15,0.5);
\draw [line width=1.2pt,color=ffcydc] (15,0.5)-- (15.5,0.5);
\draw [line width=1.2pt,color=ffcydc] (15.5,0.5)-- (15.5,1);
\draw [line width=1.2pt,color=ffcydc] (15.5,1)-- (15,1);
\draw [line width=1.2pt,color=wqwqwq] (13,2)-- (13,1.5);
\draw [line width=1.2pt,color=wqwqwq] (13,1.5)-- (13.5,1.5);
\draw [line width=1.2pt,color=wqwqwq] (13.5,1.5)-- (13.5,2);
\draw [line width=1.2pt,color=wqwqwq] (13.5,2)-- (13,2);
\draw[color=wqwqwq, line width=1.3pt] (6,-2) -- (12,-2);
\end{tikzpicture}
\caption{
Typical structure of sets of behaviors of interest.}
\label{fig:sets}
\end{figure}
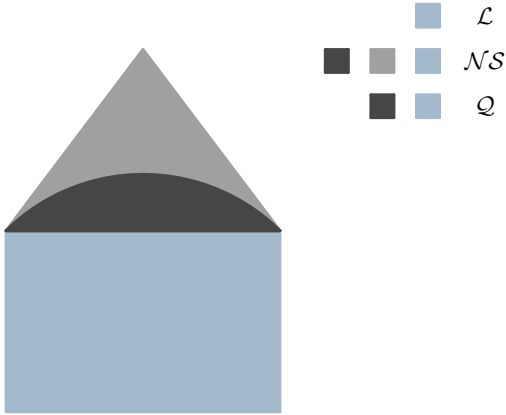

In general, the set of local behaviors $\mathcal{L}$ is a strict subset of the quantum behaviors $\mathcal{Q}$  that in turn is a strict subset of nonsignaling behaviors $\mathcal{NS}$, as shown in fig. \ref{fig:sets}. 

The local set is a polytope, and hence any local behavior can be written as a convex sum of a finite set of  extremal points. If we represent a behavior $p(a,b \vert x,y)$ as a vector $\p$ with  $\vert x\vert \vert y\vert \vert a\vert \vert b\vert$ components,  condition \eqref{LHV} can be written succinctly as 
\begin{equation}\p=A \cdot \bm{\lambda},\end{equation}
with $\bm{\lambda}$ being a probability vector over the set of variables $\lambda$, with components $\lambda_i=p(\lambda=i)$, and $A$ being a matrix indexed by $i$ and the multi-index variable $j=(x,y,a,b)$  with $A_{j,i}=\delta_{a,f_a(x,\lambda=i)}\delta_{b,f_b(y,\lambda=i)}$, where $f_a$ and $f_b$ are deterministic functions that give the values of measurements $x$ and $y$ when $\lambda=i$. Thus, checking whether $ \p$ is local amounts to a simple feasibility problem that can be written as the following linear program  \cite{Brunner2014,Chaves2015b,Fitzi2010,Abramsky2016,Rosier2017}:
\begin{eqnarray}\label{eq:LP}
\min_{\lambda \in \RR^m} & & \quad \quad \mathbf{v} \cdot \bm{\lambda}  \\ \nonumber
\st & &  \quad\p=A \cdot \bm{\lambda}  \\ \nonumber
& & \quad \lambda_i \geq 0 \\ \nonumber 
& & \quad \sum_{i}\lambda_i=1, \nonumber
\end{eqnarray}
where $\mathbf{v}$ represents an arbitrary vector with the same dimension $m=\vert x\vert^{\vert a\vert } \vert y\vert^{\vert b\vert }$ as the vector representing the  variable $\bm{\lambda}$.

Another way of testing membership in $\mathcal{L}$  is using a \emph{Bell inequality}, which provides a simple criterion for assessing whether a behavior is local or nonlocal. A Bell inequality is a linear inequality that is satisfied by all local behaviors but is violated by some nonlocal one. These inequalities correspond to hyperplanes that separate the local set from a nonlocal behavior, and thus give a sufficient condition for nonlocality.
Since the local set is a polytope, it can be characterized by a finite number of facet-defining Bell inequalities. Hence, testing membership in $\mathcal{L}$ can be done testing if the behavior satisfies all the facet-defining Bell inequalities for that scenario. Although this is equivalent to the LP formulation, it can not be done in practice because finding the set of all facet-defining Bell inequalities is an extremely hard problem  \cite{PhysRevA.97.022104}.

\section{Quantifying nonlocality of a behavior}\label{section3}

Besides identifying nonlocal distributions, Bell inequalities provide a way to quantify how nonlocal is a distribution $p$. The most common measure of nonlocality is to associate a greater numerical violation of a Bell inequality with a greater degree of nonlocality of $p$. However, this gives a nonlocality quantifier that is not {\it faithful}, since in general there are many nonlocal distributions that will not violate a given inequality. One could in principle maximize this quantity over all inequalities for that scenario, but this is in general infeasible due to the complexity of finding these inequalities.

The LP \eqref{eq:LP} can be slightly modified to define a  nonlocality quantifier  based on the trace distance \cite{PhysRevA.97.022111} between two probability distributions $\q=q(x)$ and $\p=p(x)$:
\begin{equation}
D(\q,\p)=\frac{1}{2}\sum_{x} \vert q(x) -p(x) \vert. 
\end{equation}
To quantify nonlocality we use the distance between the probability distribution generated in a Bell experiment and the closest classical probability. We are then interested in the trace distance between $q(a,b,x,y)=q(a,b \vert x,y) p(x,y)$ and $p(a,b,x,y)=p(a,b \vert x,y) p(x,y)$, where $p(x,y)$ is the probability of the inputs and that we choose to fix as the uniform distribution, that is, $p(x,y)=\frac{1}{\vert x \vert \vert y \vert}$. 

\begin{figure}[h!]
\centering
\definecolor{xdxdff}{rgb}{0.49019607843137253,0.49019607843137253,1}
\definecolor{ududff}{rgb}{0.30196078431372547,0.30196078431372547,1}
\definecolor{bcduew}{rgb}{0.7372549019607844,0.8313725490196079,0.9019607843137255}
\definecolor{qqwwzz}{rgb}{0,0.4,0.6}
\begin{tikzpicture}[scale=0.6 ,line cap=round,line join=round,>=triangle 45,x=1cm,y=1cm]
\fill[line width=2pt,color=qqwwzz,fill=qqwwzz,fill opacity=1] (5,1) -- (5,-3) -- (11.292061789564135,-3) -- (11.292061789564135,1) -- cycle;
\fill[line width=2pt,color=bcduew,fill=bcduew,fill opacity=1] (5,1) -- (11.292061789564135,0.9971334792922153) -- (9.34,4.44) -- (5.84,3.22) -- cycle;
\draw [line width=2pt] (8.16,3.08)-- (8.185723211547879,0.9985486567264298);
\begin{large}
\draw [fill=black] (8.16,3.08) circle (3pt);
\draw[color=black] (8.16,3.48) node {$\q$};
\draw [fill=black] (8.185723211547879,0.9985486567264298) circle (3pt);
\draw[color=black] (8.185723211547879,0.5985486567264298) node {$\p^*$};
\draw[color=black] (7.6,2.22702504330775) node {$d$};
\end{large}
\begin{Large}
\draw[color=black] (8.235821423467513,-1.1283849067653924) node {$\mathcal{L}$};
\draw[color=black] (9.503240193955191,1.9751661850698143)  node {$\mathcal{NS}$};
\end{Large}\end{tikzpicture}
\caption{Schematic drawing of a distribution $\q \in \mathcal{NS}$ and $d=\mathrm{NL}\left(\q\right)$,
the distance (with respect to the $\ell_1$ norm) from $\q$ to the closest local distribution $\p^* \in \mathcal{L}$.}
\label{fig:dist}
\end{figure}
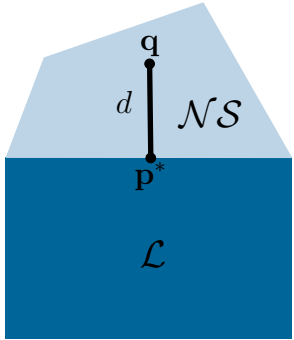

We define the quantifier $\mathrm{NL}(\q) $ for the nonlocality of behavior  $\q=q(a,b \vert x,y)$ as
\begin{eqnarray}
\label{NLtrace}
\mathrm{NL}(\q) & & =\frac{1}{\vert x \vert \vert y \vert } \min_{\p \in \mathcal{L}} \quad D(\q,\p)\\ \nonumber
& & =\frac{1}{2\vert x \vert \vert y \vert } \min_{\p \in \mathcal{L}} \sum_{a,b,x,y} \vert q(a,b \vert x,y) - p(a,b\vert x,y) \vert.
\end{eqnarray}
This is the minimum trace distance between the distribution under test and the set of local distributions. Geometrically it can be understood as how far we are from that set with respect to the $\ell_1$-norm, as shown in fig \ref{fig:dist}. 

\section{Quantifying nonlocality of a quantum state}\label{section4}

We now address the problem of quantifying nonlocality of a quantum state $\rho$. Since from $\rho$ we can generate many different nonlocal behaviors $p$, this is not a trivial problem.

One way of defining a nonlocality quantifier for quantum states is to maximize the degree of violation of a Bell inequality over all possible measurements for Alice and Bob and to associate a greater numerical violation  with a greater degree of nonlocality. This association has generated some debatable conclusions. For example, using the usual measure, the so called ``anomaly of nonlocality'' appears \cite{10.5555/2011706.2011716,PhysRevA.65.052325}. 
Consider the Collins-Gisin-Linden-Massar-Popescu  (CGLMP) inequality in the $(2,2,3)$ scenario \cite{PhysRevLett.88.040404}:

\begin{align}CGLMP &= p\left(a=b\vert 00\right) + p\left(a=b\vert 01\right) \nonumber\\
&+p\left(a=b\vert 10\right) + p\left(a=b+2\vert 11\right)\nonumber\\
&-p\left(a=b+1\vert 00\right) - p\left(a=b+2\vert 01\right) \nonumber\\
&- p\left(a=b+2\vert 10\right)
-p\left(a=b\vert 11\right) \leq 2,\end{align}
and a two qutrit system in state

\begin{equation}\label{eq:cglmpparamet}
    \ket{\psi\left(\gamma\right)}=\frac{1}{\sqrt{2+\gamma^2}} \left(\ket{00}+\gamma\ket{11}+\ket{22}\right).
    \end{equation}
For the maximally entangled state $\gamma=1$, the best choice of measurements gives $CGLMP = \frac{4(2\sqrt{3}+3)}{9} \simeq 2.873$ \cite{PhysRevLett.88.040404}. However, the authors in \cite{PhysRevA.65.052325} found that, for the very
same choice of settings, another state gives a higher violation. Specifically, the violation
$CGLMP = 1 + \sqrt{\frac{11}{3}} \simeq 2.915$ is obtained for the non-maximally entangled state
with $\gamma = \frac{\sqrt{11}-\sqrt{3}}{2} \simeq 0.792.$
This fact is known as \emph{anomaly of nonlocality}. 

Besides the non-expected feature of anomaly of nonlocality, there are other debatable issues with using the numerical value of the violation of a specific Bell inequality to quantify nonlocality. State $\rho$ having a greater numerical violation than $\sigma$ does not influence the fact that the correlations generated with both can not be explained by local realistic models. In other words, given an experimental configuration in some Bell experiment, Bell inequalities can only say whether the behavior obtained with some state is local or nonlocal.

In reference \cite{Fonseca2015}, the authors present an alternative measure to quantify nonlocality, called \emph{volume of violation}. While the previous measure takes only the settings that lead to the maximum violation, the volume of violation takes into account all the settings that produce violation of a Bell inequality. For the calculation of this new quantity, for a particular state, we make an integration in the region that leads to the violation of  a fixed Bell inequality \cite{10.5555/2011706.2011716}. In general, we can write
\begin{equation}
    V_I(\rho) = \frac{1}{V_T}\int_{\Gamma} d^n x,
\end{equation}
where $\Gamma$ is the set of measurement choices for Alice and Bob that lead to a violation of the inequality $I$ for state $\rho$ and 
\begin{equation}
    V_T = \int_{\Lambda} d^n x ,
\end{equation}
is the volume of the set $\Lambda$ of all possible measurement choices for Alice and Bob. Note that $d^n x$ will display the format that gives equal weights to any setting. Thus, state $\rho$ is more nonlocal than state $\sigma$ iff $V_I(\rho) > V_I(\sigma)$. Also, if for state $\rho$ the  volume of violation is $V_I(\rho) = 0$, we say that state $\rho$ is local with respect to the given inequality. Following the same reasoning, $V_I(\rho) = 1$ indicates that $\rho$ is maximally nonlocal with respect to that inequality. 

This measure uses  the relative volume of measurement choices that lead to violation of a particular Bell inequality to quantify nonlocality. Hence, $V_I(\rho)$ has a direct interpretation: it corresponds to the probability of violating a particular Bell inequality with state $\rho$ when the measurement configuration is chosen randomly.

The volume of violation is a measure of nonlocality for quantum states with many good properties, as already reported in reference \cite{Fonseca2015}. Nevertheless, $V_I(\rho)$ does not take into account all the measurement configurations that lead to nonlocal behaviors, but only the ones that lead to nonlocal behaviors that violate a particular Bell inequality. Except for the simplest scenario $(2,2,2)$, there are many nonequivalent families of Bell inequalities and hence $V_I(\rho)$ gives only a lower bound to the relative volume of measurement choices that lead to nonlocality.

In reference \cite{PhysRevA.96.012101} the authors consider a modification of the volume of violation, called \emph{nonlocal volume}, replacing the violation of a Bell inequality by membership of the corresponding behavior in the polytope of local correlations. Hence, the nonlocal volume  takes into account all the settings that produce a nonlocal behavior, which is in general strictly larger than the set of behaviors violating a single Bell inequality. For the calculation of this new quantity, for a particular state, we make an integration in the region $\Delta$ of the set of measurement setups that lead to  a nonlocal behavior. In general, we can write
\begin{equation}
    V(\rho) = \frac{1}{V_T}\int_{\Delta} d^n x,
\end{equation}
where $V_T$ is defined as before.
Again we consider an integration that gives equal weights to any setting. 

In this contribution  we consider another  option in which we  integrate over the set of measurement choices that give a nonlocal behavior but using a nonlocality quantifier $Q$ defined in the set of behaviors as a weight in the integral:
\begin{equation}
    V_Q(\rho) = \frac{1}{V_T}\int_{\Delta}Q(x) d^n x,
\end{equation}
where $Q(x)$ is the quantifier $Q$ for the behavior generated by the state and measurement settings $x$ and $\Delta$ and $V_T$ are defined as before.
We consider only \emph{faithful} quantifiers: $Q(\p) > 0$ iff $\p \notin \mathcal{L}$.

This new quantifier, that we call \emph{$Q$-weighted nonlocal volume},  can be interpreted in a similar way. As the nonlocal volume, it  takes into account all the settings that produce a nonlocal behavior for state $\rho$, but it sums with a higher weight the behaviors that are more nonlocal according to the quantifier $Q$. 

In particular, we are interested in the nonlocality quantifier for states obtained when we choose $Q=NL$: 

\begin{equation}
    V_{NL}(\rho) = \frac{1}{V_T}\int_{\Delta}NL(x) d^n x,
\end{equation}
where $NL(x)$ is the trace distance  for the behavior generated by the state and measurement settings $x$ and $\Delta$ and $V_T$ are defined as before.
We call this quantifier trace-weighted nonlocal volume.

\section{Properties}\label{section5}
 In reference \cite{lipinska2018towards}, the authors show that the nonlocal volume is invariant under local unitaries and that it is strictly positive for entangled bipartite states. The proofs can be slightly modified to show that these properties are also true for $V_Q$. 

\begin{thm}
$V_Q$ is invariant under local unitaries.
\end{thm}

\begin{proof}
Let $\rho'=U_1\otimes U_2 \rho U_1^\dagger\otimes U_2^\dagger$ where $U_1$ and $U_2$ are local unitaries for subsystems $1$ and $2$, respectively. The behavior generated with measurements $\left\{M_i\right\}$ and $\left\{N_i\right\}$ and state $\rho$ is equal to the behavior generated with measurements $\left\{U_1^\dagger M_iU_1\right\}$ and $\left\{U_2^\dagger N_iU_2\right\}$ and state $\rho'$. Hence, the sets $\Delta$ are the same for $\rho$ and $\rho'$, which implies $V_Q(\rho)=V_Q(\rho')$.
\end{proof}

\begin{thm}
If $Q$ is a continuous function, for all pure bipartite entangled states, in a scenario with at least two choices of two-outcome measurements, $V_Q$ is strictly positive.
That is, $V_Q\left(\ket{\psi}\right)=0$ if and only if $\ket{\psi}$ is a product state.

\end{thm}

\begin{proof}
Since $\ket{\psi}$ is entangled, we know from \cite{gisin1991bell}, that there exist measurements $\left\{M_i\right\}$, $\left\{N_j\right\}$  in the simplest scenario $(2,2,2)$ such that the corresponding behavior is nonlocal. By continuity of the probabilities $p\left(ab\vert xy\right)$ as a function of the measurements, and continuity of $Q$ as a function of $\p$, there is a ball around $\left\{M_i\right\}, \left\{N_j\right\}$  such that, for any choice of measurements inside this ball, $Q$ is strictly positive. Since we are integrating a strictly positive function over a set that is of measure larger than zero, this implies that    $V_Q\left(\ket{\psi}\right)>0$.

On the other hand, if $\ket{\psi}$ is separable, every behavior generated with $\ket{\psi}$ is local, and hence $Q(x)=0$ for every $x$ and $V_Q\left(\ket{\psi}\right)=0$.
\end{proof}

It is also  easy to see that the $Q$-weighted nonlocal volume is always smaller than the non-weighted version  for any faithful quantifier $Q$ that is normalized such that $0\leq Q \leq 1$.
In fact, if $Q$ is faithful, we can write
\begin{equation}
    V_{Q}(\rho) = \frac{1}{V_T}\int_{\Lambda}Q(x) d^n x,
\end{equation}
and
\begin{equation}
    V(\rho) = \frac{1}{V_T}\int_{\Lambda}\chi(x) d^n x,
\end{equation}
where $\chi$ is the characteristic function of the nonlocal set, that is:
\begin{equation}
    \chi(\p)=\left\{\begin{array}{cc}
         1,&  \ \mbox{if} \  \p \notin \mathcal{L},\\
         0, &  \ \mbox{if} \  \p \in \mathcal{L}.\end{array}\right.
\end{equation}
If $\p$ is local and $Q$ is faithful, $Q(\p)=\chi(\p)=0$. If $\p$ is nonlocal, 
$Q(\p)\leq 1=\chi(\p)$, which implies that $Q(\p)\leq \chi(\p)$ for all $\p$ and hence $V_Q(\rho) \leq V(\rho)$ for all $\rho$.

\begin{figure}[ht]
    \centering
    \includegraphics[width=0.5\textwidth]{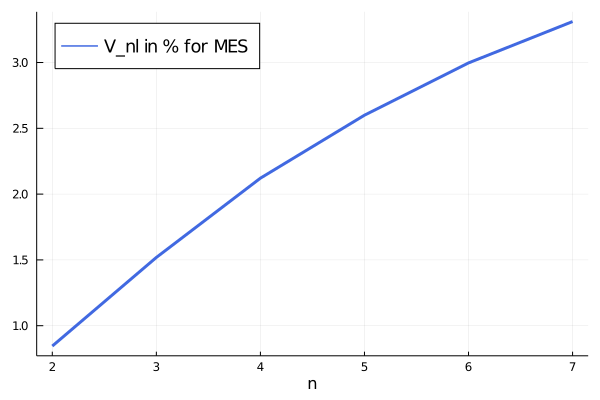}
    \caption{Trace-weighted nonlocal volume in percentage for the maximally entangled two-qubit state as function of the number of measurements detained by each part in the Bell scenario, i.e scenarios of the type $(2,n,2)$.}
    \label{fig:ntoinf}
\end{figure}

Reference \cite{lipinska2018towards} also shows that the nonlocal volume  goes to $1$ in the limit where both parties have an infinite number of measurements. The proof in this case can not be modified to show that this is also true for $V_Q$. Numerical results in the figure \ref{fig:ntoinf} indicate that if we choose $Q=NL$, $V_{NL}$ seems to increase monotonically with the number of measurements $n$, which would be the desired behavior. However, it is not possible to claim its limit as $n$ goes to infinity is 1 based on this evidence only. Since $V_Q$ is always smaller than the nonlocal volume, we can not discard the possibility that $V_Q$ goes to $a$ as $n$ goes to infinity, with $0<a<1$.  Whether or not this property holds for $V_{NL}$ is an open problem.

\section{Revisiting simple scenarios}\label{section6}
In this section we test the quantifier $V_{NL}$, analysing its performance when compared to previous  results in the literature. Behaviors are decided to be nonlocal in our simulations for trace distances with magnitude smaller than $10^{-8}$.

\begin{figure}[ht]
    \begin{subfigure}{.49\textwidth}
     \centering
     \includegraphics[width=1\linewidth]{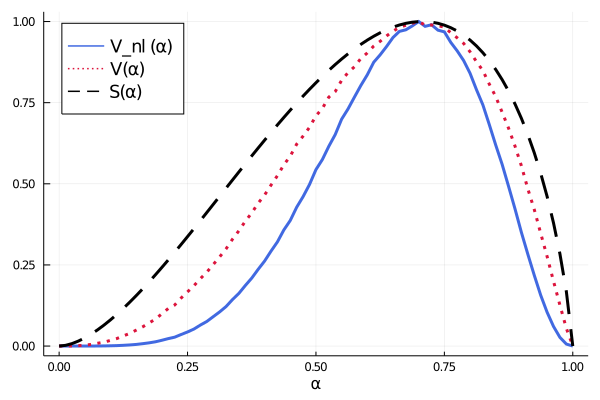}
     \caption{}
     \label{subfig:chsh_comp}
    \end{subfigure}
    \begin{subfigure}{.49\textwidth}
     \centering
     \includegraphics[width=1\linewidth]{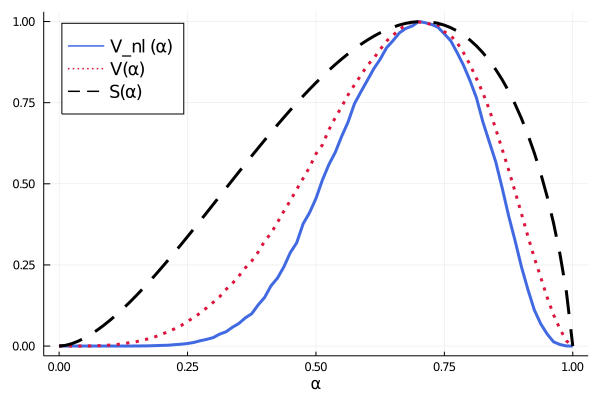}
     \caption{}
     \label{subfig:3322_comp}
    \end{subfigure}
    \caption{In blue our quantifier lies below its non-weighted version in red, as expected, while maintaining the point of maximum, which coincides with the maximum of the dashed curve representing the entanglement entropy. In (a) for the CHSH scenario and in (b) for the 3322 scenario.}
    \label{fig:comparison}
\end{figure}

To start, we check the effect of the weight on the nonlocal volume for the simplest scenario, the $(2,2,2)$ scenario, also known as the CHSH scenario. The local set in this case has a simple structure, since there is only one family of Bell inequalities, and we expect that the plots for the entropy of entanglement, the nonlocal volume, and the $Q$-weighted nonlocal volume all have the same comportment if $Q$ is a continuous faithful nonlocality quantifier for behaviors. To test that we consider the family of states parametrized by:
\begin{equation}
    \ket{\psi(\alpha)}=\alpha\ket{00}+\sqrt{1-\alpha^2}\ket{11}\label{eq:parqubit}
\end{equation}
and plot the nonlocal volume (red) and the trace-weighted nonlocal volume (blue) in figure \ref{subfig:chsh_comp}. 
Both for the nonlocal volume and the trace-weighted nonlocal volume the maximum nonlocality is attained at the maximally entangled state. We can also see that the weighted version is smaller or equal the non-weighted version considering a normalization using the maximum value for each curve. 

The next interesting scenario, the $(2,3,2)$ scenario,  is not as trivial as the CHSH scenario. In figure \ref{subfig:3322_comp}   the  weighted and the non-weighted version are compared for the qubits states in the same parametrization \eqref{eq:parqubit}.

\begin{figure}[ht]
    \centering
    \includegraphics[width=0.5\textwidth]{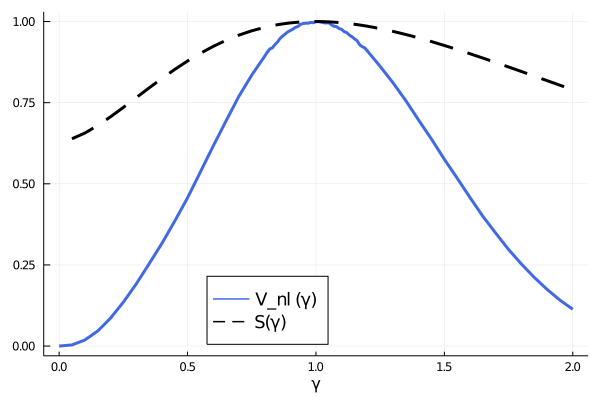}
    \caption{The maximum of the trace-weighted nonlocal volume (in blue) is reached at the state exhibiting the maximum entanglement entropy (in black).}
    \label{fig:qutrits_eevnl}
\end{figure}

We also compared $V_{NL}$ and the entropy of entanglement for the family of states given in \eqref{eq:cglmpparamet}, see figure \ref{fig:qutrits_eevnl}. That is, for the scenario of qutrits. For this family of states, we see that the maximum nonlocality is achieved by the maximally entangled state, as in the previous scenarios with qubits. The same behavior is observed for the non-weighted version, as shown in reference \cite{PhysRevA.96.012101}. But in general $V_{NL}$ is not a monotonic function of the entropy of entanglement. This property is known as \emph{weak anomaly of nonlocality}.

 \section{Weak Anomaly of Nonlocality}\label{section7}

\begin{figure}[ht]
    \begin{subfigure}{.49\textwidth}
     \centering
     \includegraphics[width=1\linewidth]{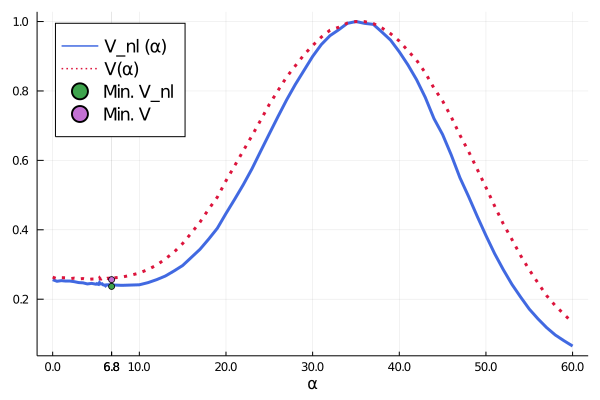}
     \caption{}
     \label{subfig:anomaly1}
    \end{subfigure}
    \begin{subfigure}{.49\textwidth}
     \centering
     \includegraphics[width=1\linewidth]{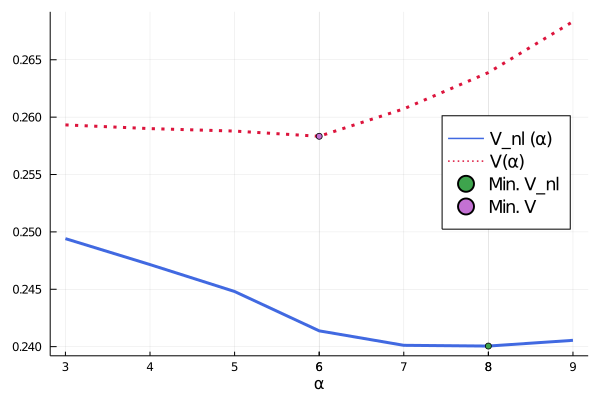}
     \caption{}
     \label{subfig:anomaly2}
    \end{subfigure}
    \caption{Weak anomaly from the perspective of the trace-weighted nonlocal volume. In (a) we see how the weight compresses the normalized curve slightly, while in (b) a zoomed view of the region containing the local minimum shows that each quantifier attains the minimum at a different angle (bold ticks).}
    \label{fig:weakanomaly}
\end{figure}

Still in the same system of qutrits, for the GHZ($\alpha$) states parametrized as:
\begin{equation}\label{eq:ghzalpha}
    \text{GHZ}(\alpha) = \sin(\alpha)|00\rangle+\frac{\cos(\alpha)}{\sqrt{2}}\left(|11\rangle+|22\rangle\right)\ ,
\end{equation}
nonlocality does not increase monotonically with $\alpha$. As in the previous cases, it reaches the peak at the maximally entangled state, but it has a local minimum around 6 degrees, similar to what is observed in the non-weighted version. Figure \ref{subfig:anomaly1} shows the weighted and non-weighted versions of nonlocal volume as a function of $\alpha$; figure \ref{subfig:anomaly2} gives a zoom on what is happening very close to the local minimum, emphasizing the different behavior of the quantifiers.

It is interesting to observe that the local minima are reached for different values of $\alpha$ for the weighted and non-weighted version. This shows that the weak anomaly is not an intrinsic characteristic of the scenario, but is strongly dependent on the choice of quantifier.

To investigate further how entanglement and nonlocality can behave differently, we study how these quantities change when we go from a lower rank state to a full rank state. We first show that entanglement never decreases in this situation.

\begin{thm}
Entanglement always increases when making a contiuous transition from a lower rank state $\ket{\psi}$ to a full rank state $\ket{\psi'}$. 
\end{thm}

\begin{proof} Let $\ket{\psi} \in \mathcal{H}=\mathcal{H}^{(A)}\otimes\mathcal{H}^{(B)}$, $\dim\mathcal{H}=D$. Suppose that rank of $\ket{\psi}$ is smaller than $\sqrt{D}$, so that its Schmidt decomposition reads:
\begin{equation}
    \ket{\psi}=\sum^{d-1}_{i=0}\alpha_i \ket{ii}, 
\end{equation}
where $\sum^{d-1}_{i=0}|\alpha_i|^2 = 1$ and $d<\sqrt{D}$, leading to the entropy of entanglement:
\begin{equation}
    E=-\sum^{d-1}_{i=0}|\alpha_i|^2 \log_2 |\alpha_i|^2 \text{Ebits}.
\end{equation}
Consider now another state $\ket{\psi'}$ which is full rank, $rank(\ket{\psi'})=\sqrt{D}$
\begin{equation}
    \ket{\psi'}=\frac{1}{N}(\delta\ket{\phi}+\ket{\psi}),
\end{equation}
where $\ket{\phi}=\sum^{\sqrt{D}}_{i=d}\beta_i \ket{ii}$, $\sum^{\sqrt{D}}_{i=d}|\beta_i|^2 =1$, $\delta\in\mathbb{C}$ and $N$ is a normalization factor. Using that $\braket{\phi}{\psi}=0$
$$\ket{\psi'}=\frac{1}{\sqrt{1+|\delta|^2}}(\delta\ket{\phi}+\ket{\psi}),$$
whose entropy of entanglement is given by:
\begin{eqnarray*}
E'=&-&\frac{|\delta|^2}{1+|\delta|^2}\sum^{\sqrt{D}}_{i=d}|\beta_i|^2\log\left(\frac{|\delta|^2|\beta_i|^2}{1+|\delta|^2}\right)\\
&-&\frac{1}{1+|\delta|^2}\sum^{d-1}_{i=0}|\alpha_i|^2\log\left(\frac{|\alpha_i|^2}{1+|\delta|^2}\right).
\end{eqnarray*}
The first sum can be written as
\begin{eqnarray*}
& &=-\sum^{\sqrt{D}}_{i=d}|\beta_i|^2\left[\log|\delta|^2+\log|\beta_i|^2-\log(1+|\delta|^2)\right]\\
& &=-\log|\delta|^2+E_{\phi}+\log(1+|\delta|^2).
\end{eqnarray*}
The second sum becomes
\begin{eqnarray*}
& &=-\sum^{d-1}_{i=0}|\alpha_i|^2\left[\log|\alpha_i|^2 - \log(1+|\delta|^2)\right]\\
& &=E+\log(1+|\delta|^2).
\end{eqnarray*}
Thus
\begin{eqnarray*}
=&&\frac{1}{1+|\delta|^2}\biggl[-|\delta|^2\log|\delta|^2+|\delta|^2 E_{\phi}\\
&+& |\delta|^2\log(1+|\delta|^2)+E+\log(1+|\delta|^2) \biggr].
\end{eqnarray*}
Now we take $|\delta|^2\ll 1$
\begin{eqnarray*}
E'\to (1-|\delta|^2)\biggl[&-&|\delta|^2\log|\delta|^2 +|\delta|^2 E_{\phi}\\&+&|\delta|^4+E+|\delta|^2\biggr].
\end{eqnarray*}
Since $-\frac{|\delta|^2\log|\delta|^2}{|\delta|^2} \to \infty$ when $\delta\to 0$, the leading order expression is:
\begin{equation}
E' \sim (1-|\delta|^2)\left[-|\delta|^2\log|\delta|^2 +E \right]
\end{equation}
So, entanglement always increases when we make a continuous change form a lower rank state $\ket{\psi}$ to a full rank state $\ket{\psi'}$. In addition, the leading order entanglement gain is independent of the ``direction'' by which the new subspace in penetrated (independent of $\ket{\phi}$ and of $E_{\phi}$):
\begin{equation}
    \Delta E=E'-E=-|\delta|^2\log|\delta|^2 >0
\end{equation}
for $|\delta|^2$ small. 
\end{proof}

\begin{figure}[ht]
    \begin{subfigure}{.49\textwidth}
     \centering
     \includegraphics[width=1\linewidth]{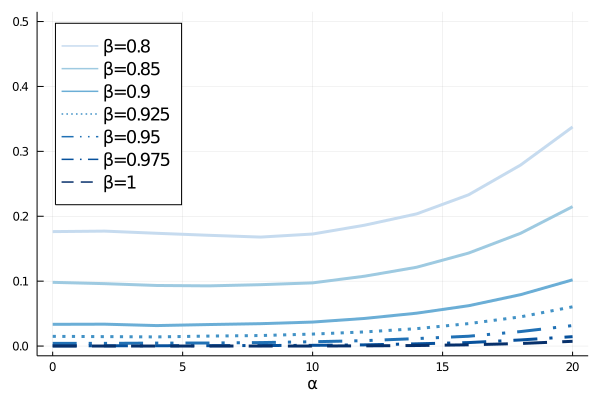}
     \caption{}
     \label{subfig:wa_rank}
    \end{subfigure}
    \begin{subfigure}{.49\textwidth}
     \centering
     \includegraphics[width=1\linewidth]{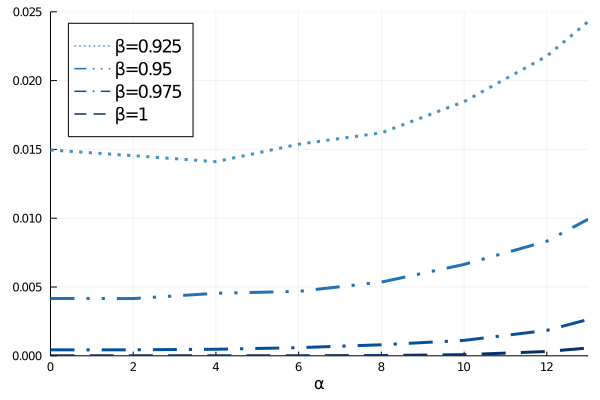}
     \caption{}
     \label{subfig:wa_rank_zoom}
    \end{subfigure}
    \caption{Trace-weighted nonlocal volume for different families of states in function of $\alpha$; a family in this case is labeled by the value of its $\beta$ parameter in $\text{GHZ}(\alpha,\beta)$. Lighter curves stand for greater entanglement among states $|11\rangle$ and $|22\rangle$.}
    \label{fig:wa}
\end{figure}

On the other hand, nonlocality can behave very differently when changing from a low rank to a full rank state. Figure \ref{fig:wa} shows  an investigation within regions of states with partial entanglement between states $|11\rangle$ and $|22\rangle$ in the GHZ state with an extra parameter:
\begin{eqnarray}
 \text{GHZ}(\alpha,\beta)&=&\sin(\alpha)|00\rangle \nonumber\\
 &+&\frac{\cos(\alpha)}{\sqrt{2}}\left(\beta|11\rangle+\sqrt{1-\beta^2}|22\rangle\right) .\nonumber\\
 \end{eqnarray}

Notice in \ref{subfig:wa_rank} that the minimum of the anomaly moves towards zero degree as we increase the value of $\beta$, i.e. as we consider a starting point closer to a rank-1 state ($|11\rangle$). This suggests that the minimum not only depends on the quantifier for behaviors that we extend to quantify nonlocality of states, but also on the region of parameters at which we are looking at. Around the neighborhood of states with different ranks the anomaly shows its unsteadiness.
 
 Figure \ref{subfig:wa_rank_zoom} reinforce this observation, while also showing that there is a family of states (with $\beta=0.975$) for which the quantifier is monotonic as a function of $\alpha$. Increasing that value would lead to other monotonic curves too. 
 
 This shows that relation between entanglement and nonlocality is not simple, depending on many aspects of the scenario and the tools we use to quantify these properties. Although this may appear counter intuitive at first sight, since nonlocality is a consequence of entanglement, it is already known that entanglement and nonlocality are indeed different resources, and the fact that nonlocality in not a monotonic function of entanglement is another feature that supports this claim.

\section{Discussion}\label{section8}
We have seen an alternative way for quantifying nonlocality of states based on Bell nonlocality of behavior. The key difference from preceding candidates was the introduction of a quantifier of nonlocality to weight each contribution from behavior in the nonlocal volume.

A new degree of freedom has then been brought to the topic. Here we explored it by considering the simplest possibility among the ones at disposal in the literature when the question is how to compare two quantum probability distributions: the trace distance. We proved that this quantifier has several good properties, including its formulation in terms of a linear programming, but our simulations have shown that it is not a monotonic function of entangleme. More specifically, the weak anomaly of nonlocality persists. Nevertheless, the local minimum for nonlocality with the trace-weighted nonlocal volume occurs  in a different state than the minimum for the non-weighted version, showing that the weak anomaly is not an intrinsic characteristic of the scenario, but is  dependent of the choice of quantifier. We conjecture that this non-monotonicity, despite the coincidence of the maxima, is a unavoidable manifestation of the intrinsic inequivalence between entanglement and nonlocality.  It is a topic to further investigation how the behavior of the weighted version would be for different quantifiers $Q$ and whether we can prove that every nonlocality quantifier for states  exhibits some kind of anomaly. Moreover, its robustness against noisy systems also remains to be tested.

It should be noted, however, that in the integral defining the nonlocal volume we considered projective measurements only. The next natural generalization would be to consider the set of  POVM's, but two difficulties appear. First, there is no natural definition of uniform measure for POVM's, which makes the quantifier strongly dependent on the choice of measure we choose to sample the measurements. Second, numerical results show that the probability of finding a nonlocal behavior when sampling over the set of POVM's is very small.  In fact, in a first attempt we have made using Neumark's dilation theorem and the Haar measure in the set of unitaries in the larger Hilbert space, typically a nonlocal behavior is observed in one out of ten million settings for the $(2,2,2)$ scenario (even for the maximally entangled state), a  simulation which, in addition, takes much more time to be finished. 

This general idea has come with the seeming necessity for refining known quantifiers in order to extinguish such anomalies against the monotonic relation expected between nonlocality and entanglement of states. An interesting venue for further research is to find out a quantifier of nonlocality for behavior that implies monotonicity of the weighted nonlocal volume and/or presents the desired property at least for projective measurements, and preferably is computationally friendly. However, we conjecture that this is not possible. If no anomaly remains, that is, if nonlocality increases monotonically with entanglement, we would conclude that nonlocality and entanglement are equivalent resources, which we know not to be true. Moreover, we know that different nonlocality quantifiers give different orderings in the set of behaviors, which makes   nonlocality heavily dependent on the function we use to quantify it.

\clearpage
\bibliographystyle{alpha}
\bibliography{main}

\end{document}